\documentclass[aps,showpacs,preprint,footinbib,preprintnumbers]{revtex4-1}

\usepackage{graphicx}
\usepackage{epstopdf}
\usepackage{amsmath,amssymb}
\usepackage{booktabs}
\usepackage{mathrsfs}
\usepackage{multirow}

\newcommand{\be}{\begin{eqnarray}}
\newcommand{\ee}{\end{eqnarray}}

\newcommand{\vslash}{{v\hspace{-5.4pt}/}}

\usepackage[normalem]{ulem}  
\usepackage[dvips]{color} 
\renewcommand\sout{\bgroup \color{red} \ULdepth=-.5ex \ULset}

\begin{document}

\title{Hadronic molecules for charmed and bottom baryons near thresholds}

\author{Yasuhiro Yamaguchi$^1$}
\author{Shunsuke Ohkoda$^1$}
\author{Shigehiro Yasui$^2$}
\author{Atsushi Hosaka$^1$}
\affiliation{$^1$Research Center for Nuclear Physics (RCNP), 
Osaka University, Ibaraki, Osaka, 567-0047, Japan}
\affiliation{$^2$KEK Theory Center, Institute of Particle and Nuclear
Studies, High Energy Accelerator Research Organization, 1-1, Oho,
Ibaraki, 305-0801, Japan}

\date{\today}

\begin{abstract}
We study hadronic molecules formed by a heavy meson and a nucleon,
 $DN$ and $D^{\ast}N$ ($\bar{B}N$ and $\bar{B}^{\ast}N$) systems.
Respecting the heavy quark symmetry and chiral symmetry,
we consider the $DN$-$D^\ast N$ ($\bar{B}N$-$\bar{B}^\ast N$) mixing
induced by the one boson exchange potential including the tensor force.
We find many bound and resonant states with $J^P=1/2^\pm$, $3/2^\pm$,
 $5/2^\pm$ and $7/2^-$
in isospin singlet channels,
while only a few resonant states with $J^P=1/2^-$ in isospin triplet channels.
The analysis of $DN$ and $D^{\ast}N$ ($\bar{B}N$ and $\bar{B}^{\ast}N$) molecules will be useful to study mass spectra of excited charmed (bottom)
baryons with large angular momenta, when their masses are close to the $DN$ and $D^\ast N$ ($\bar{B}N$ and $\bar{B}^{\ast}N$) thresholds.

\end{abstract}
\pacs{12.39.Hg, 14.20.Lq, 14.20.Mr, 14.20.Pt}
\maketitle

\section{Introduction}
Recently, many exotic hadrons observed in experiments are attracting our interest.
They are expected to have unexpected structures, because their properties, such as quantum numbers, masses, and decay patterns, cannot be explained by
the ordinary picture of hadrons, baryons as $qqq$ and mesons as $q\bar{q}$.
As one of new aspects of these structures,
the possibility of multiquarks or hadronic molecules have been
extensively discussed. 
It is considered that, particularly near thresholds,
hadronic molecules such as loosely bound or resonant states of
meson-meson or meson-baryon 
emerge as one of hadronic configurations realized in nature.
As typical candidates,
the meson-meson molecules for $X(3872)$ and $Z_b$ have been studied 
in the charm and bottom quark regions~\cite{Choi:2003ue,Voloshin:2007dx,Swanson:2006st,Brambilla:2010cs,Belle:2011aa,Adachi:2012im,Voloshin:2011qa,Bondar:2011ev}.

The hadronic molecules as bound or resonant states formed by $D$ or $D^{\ast}$ meson ($\bar{B}$ or $\bar{B}^{\ast}$ meson) and a
nucleon ($N$) are also interesting.
Since the $DN$ and $D^{\ast}N$ ($\bar{B}N$ and $\bar{B}^{\ast}N$) molecules contain ordinary flavor structures of three quarks,
they are intimately related to the ordinary heavy
baryons, $\Lambda_{\rm c}$, $\Sigma_{\rm c}$, $\Sigma_{\rm c}^{\ast}$
and their excited states for the charm sector, and $\Lambda_{\rm b}$,
$\Sigma_{\rm b}$, $\Sigma_{\rm b}^{\ast}$ and their excited states for
the bottom sector~\cite{comment1}.
These excited states have been studied in the quark model extensively~\cite{Copley:1979wj,Roberts:2007ni}.
However, near the $DN$ and $D^{\ast}N$ ($\bar{B}N$ and $\bar{B}^{\ast}N$) thresholds,
it is expected that the wave functions of the quark model states
have a large overlap with
the component of $DN$ and $D^{\ast}N$ ($\bar{B}N$ and $\bar{B}^{\ast}N$), and
the properties of such baryon states 
are strongly affected by $DN$ and $D^{\ast}N$ ($\bar{B}N$ and
$\bar{B}^{\ast}N$) states.
There would be even such states that are dominated by the molecular components.

In the $DN$ and $D^{\ast}N$ ($\bar{B}N$ and $\bar{B}^{\ast}N$) systems, there are two important symmetries; the heavy quark symmetry and chiral symmetry.
The heavy quark symmetry manifests in 
mass degeneracy of heavy pseudoscalar mesons $P=D$, $\bar{B}$ and vector mesons
$P^\ast =D^\ast$, $\bar{B}^\ast$ in the heavy quark mass limit~\cite{Isgur:1989vq,Isgur:1991wq}.
Indeed, the mass splitting between a $D$ ($\bar{B}$) meson and a $D^\ast$ ($\bar{B}^\ast$) meson is small;
140 MeV for $D$ and $D^\ast$ mesons and 46 MeV for $\bar{B}$ and $\bar{B}^\ast$ mesons.
The small mass splittings in the heavy flavor sectors should be compared
with the large mass splittings, $\sim 400$ MeV for $\bar{K}$ and
$\bar{K}^{\ast}$ mesons, 
and $\sim 600$ MeV for $\pi$ and $\rho$ mesons, in the light flavor sectors.
Because of this, the heavy quark symmetry introduces the mixing of $PN$ and $P^{\ast}N$ states,
where both $P$ and $P^{\ast}$ mesons are considered on the same footing as fundamental degrees of freedom in the dynamics~\cite{comment2}.
Then, physical states are given as a superposition of $PN$ and $P^{\ast}N$ states.
For convenience, in the following, we denote $P^{(\ast)}$ to stand for $P$ or $P^{\ast}$.
Thanks to the mixing of states in $P^{(\ast)}N$ systems, 
a long range force between a $P^{(\ast)}$ meson and a nucleon $N$ is
supplied by one pion exchange 
with the same coupling strengths for
$PP^\ast \pi$ and $P^\ast P^\ast \pi$ vertices~\cite{Yasui:2009bz,Yamaguchi:2011xb,Yamaguchi:2011qw}.

It is interesting that the pion exchange can exist in $P^{(\ast)}N$ systems.
As a matter of fact, it is known that the pion exchange is important for
the binding of atomic nuclei~\cite{Ikeda:2010aq}.
There the tensor force of the pion exchange that mixes channels with different
angular momenta, i.e. $L$ and $L\pm 2$, yields a strong attraction 
to generate a rich structure of bound as well as resonant states.
Because the pion exchange is a long range force, it becomes more
significant for states with larger orbital angular momenta.
In fact, it was shown that, in the nucleon-nucleon scattering,
the phase shifts in the channels with large orbital angular momenta are reproduced
almost only by the pion exchange potential~\cite{Kaiser:1997mw,Kaiser:1998wa}.
The importance of the pion exchange was also discussed in $N\bar{N}$ systems.
It was investigated that the properties of bound and resonant states in $N\bar{N}$ systems are dramatically changed,
if the tensor force is switched off~\cite{Phillips:1967zza,Dover:1978br}.
In our previous studies, 
it was also shown that the pion exchange potential played a significant
role for hadronic molecules such as $\bar{D}N$ and $BN$
baryons~\cite{Yasui:2009bz,Yamaguchi:2011xb,Yamaguchi:2011qw}, 
$D\bar{D}$ ($B\bar{B}$) mesons (e.g. $Z_b$ meson)~\cite{Ohkoda:2011vj}
and $DD$ ($BB$) mesons~\cite{Ohkoda:2012hv}.

In reality, however, 
the hadronic molecules of $P^{(\ast)}N$ do not necessarily correspond to the observed states, 
because there should be couplings not only to three quark states but
also to other meson-baryon states such as $\pi\Lambda_{\rm c}$, $\pi\Sigma_{\rm c}$ and
$\pi\Sigma^\ast_{\rm c}$
($\pi\Lambda_{\rm b}$, $\pi\Sigma_{\rm b}$ and $\pi\Sigma^\ast_{\rm b}$), and so on.
However, we may expect that such couplings are small for $DN$ and $\bar{B}N$ baryons near the thresholds.  
The reasons are that the wave functions of hadronic molecules are
spatially large as compared to the conventional three quark states, 
and that the transitions, e.g. from $DN$ ($\bar{B}N$) to $\pi\Sigma_{\rm c}$ ($\pi\Sigma_{\rm b}$),
are suppressed by a heavy quark exchange.
From those points of view, in the present discussion, we focus on the role of the $D^{(\ast)}N$ and
$\bar{B}^{(\ast)}N$ sectors and study the bound or resonant states generated by
$D^{(\ast)}N$ and $\bar{B}^{(\ast)}N$.
Recently, there have been several studies of baryon states as
$D^{(\ast)}N$ ($\bar{B}^{(\ast)}N$)
~\cite{Mizutani:2006vq,GarciaRecio:2008dp,Romanets:2012hm,GarciaRecio:2012db,Hofmann:2005sw,Hofmann:2006qx,Haidenbauer:2010ch,He:2010zq,Zhang:2012xx,Zhang:2012jk}
and $\bar{D}^{(\ast)}\Sigma_{\rm c}$ ($B^{(\ast)}\Sigma_{\rm b}$)~\cite{Wu:2010jy,Wu:2010vk,Wu:2010rv} molecules.
The present study is different from them in that we focus on the states
near the thresholds and with a large angular momentum.

This paper is organized as follows. In Sec.~\ref{int.}, we
briefly describe the interaction between a $D^{(\ast)}$ or $\bar{B}^{(\ast)}$ meson and a nucleon.
In Sec.~\ref{result}, we solve the coupled channel
Schr\"odinger equations numerically for bound and resonant states.
We discuss the results of the $D^{(\ast)}N$ and $\bar{B}^{(\ast)}N$ molecule in Sec.~\ref{Discussion}, and
 summarize the present work in the final section.

\section{Interactions}\label{int.}

Following our previous studies~\cite{Yasui:2009bz,Yamaguchi:2011xb,Yamaguchi:2011qw},
we consider $\pi$, $\rho$ and $\omega$ exchange
potentials between $P^{(\ast)}$ and $N$.
The interaction Lagrangians for heavy meson vertices are given by 
the heavy quark symmetry and chiral symmetry~\cite{Manohar:2000dt,Casalbuoni},
\begin{align}
{\cal L}_{\pi HH} &=   ig_\pi \mbox{Tr} \left[
H_b\gamma_\mu\gamma_5 A^\mu_{ba}\bar{H}_a \right]   \, , 
\label{LpiHH}\\
{\cal L}_{v HH} &=  -i\beta\mbox{Tr} \left[ H_b
v^\mu(\rho_\mu)_{ba}\bar{H}_a \right]
+i\lambda\mbox{Tr} \left[
H_b\sigma^{\mu\nu}F_{\mu\nu}(\rho)_{ba}\bar{H}_a \right] \, , 
\label{LvHH}
\end{align}
where the subscripts $\pi$ and $v$ are for the pion and vector ($\rho$ and $\omega$) meson
interactions. 
The heavy meson field $H$ is defined by
$H_a  =
\frac{1+\vslash}{2}\left[P^\ast_{a\,\mu}\gamma^\mu-P_a\gamma_5\right]$
with the four-velocity $v^\mu$ of a heavy meson,
where the subscript $a$ is for light flavors, $u$, $d$.
The conjugate field is given by
$\bar H_a = \gamma_0 H^\dagger_a \gamma_0$.
The axial current of light flavors is written by
$A^\mu=\frac{1}{2}\left(\xi^\dagger\partial^\mu
 \xi-\xi\partial^\mu \xi^\dagger\right)$
with $\xi=\exp{(i\hat{\pi}/f_\pi)}$ and the pion decay
constant $f_\pi=132$ MeV. The pion field is defined by
\begin{align}
 \hat{\pi}&=\left(
\begin{array}{cc}
\displaystyle \frac{\pi^0}{\sqrt{2}}&\pi^+ \\
 \pi^- &\displaystyle -\frac{\pi^0}{\sqrt{2}}
\end{array}
\right),
\end{align}
and the vector meson field by
\begin{align}
 \rho^\mu&=i\frac{g_V}{\sqrt{2}}\hat{\rho}_\mu \, , \\
 \hat{\rho}_\mu&=\left(
\begin{array}{cc}
 \displaystyle \frac{\rho^0}{\sqrt{2}}&\rho^+ \\
 \rho^- &\displaystyle -\frac{\rho^0}{\sqrt{2}}
\end{array}
\right)_\mu \, ,
\end{align}
where $g_V \simeq 5.8$ corresponds to the gauge coupling constant from hidden local
symmetry~\cite{Bando:1987br}.
The vector meson field tensor is written by
$F_{\mu\nu}(\rho)=\partial_\mu \rho_\nu-\partial_\nu \rho_\mu+[\rho_\mu,\rho_\nu]$.
The coupling constants $g_\pi$, $\beta$ and $\lambda$ are summarized in
Table~\ref{constants}.
These coupling constants are essentially the same as those given in
Refs.~\cite{Yamaguchi:2011xb,Yamaguchi:2011qw}, except for the signs of
$g_{\pi}$, $\beta$ and $\lambda$ for vertices of $\pi$ and $\omega$,
which are reversed due to G-parity transformation between $D^{(\ast)}$ ($\bar{B}^{(\ast)}$) and
$\bar{D}^{(\ast)}$ ($B^{(\ast)}$).
In Eqs.~\eqref{LpiHH} and \eqref{LvHH}, we consider the static approximation $v^\mu=(1,\vec{0})$.

 \begin{table}[t]
  \caption{Masses and coupling constants of mesons
  $\alpha=\pi$, $\rho$, $\omega$ in Refs.~\cite{Yamaguchi:2011xb,Yamaguchi:2011qw}.}
  \label{constants}
  \begin{tabular}{lcccccc}
   \hline
   Meson&$m_\alpha$ &$g_\pi$ &$\beta$ &$\lambda$ [GeV$^{-1}$]
   &$g^2_{\alpha NN}/4\pi$ &$\kappa$ \\
   \hline
   $\pi$&137.27&$-0.59$&---&---&13.6&--- \\
   $\rho$&769.9&---&0.9&0.59&0.84&6.1 \\
   $\omega$&781.94&---&$-0.9$&$-0.59$&20.0&0.0 \\
   \hline
  \end{tabular}
 \end{table}

For vertices of $\pi$, $\rho$ and $\omega$ mesons to a nucleon, we employ the Bonn model~\cite{machleidt} as
\begin{align}
 {\cal L}_{\pi NN} &= \sqrt{2} ig_{\pi
NN}\bar{N}\gamma_5 \hat{\pi} N \, , \label{LpiNN} \\
{\cal L}_{vNN} &= \sqrt{2} g_{vNN}\left[\bar{N} \gamma_\mu \hat{\rho}^\mu N 
+\frac{\kappa}{2m_N}\bar{N} \sigma_{\mu\nu} \partial^\nu \hat{\rho}^\mu
N \right] \, , \label{LvNN}
\end{align}
where $N=(p,n)^T$ is the nucleon field.
The coupling constants for nucleons are summarized in Table~\ref{constants}.

When we consider the $\pi$, $\rho$ and $\omega$ exchange potentials between $P^{(\ast)}$ meson and nucleon,
we take into account internal structure of hadrons by introducing the
monopole-type form factors at each vertex; 
\begin{align}
 &F_\alpha(\Lambda,\vec{q}\,)=\frac{\Lambda^2-m^2_\alpha}{\Lambda^2+|\vec{q}\,^2|},
 \label{formfactor}
\end{align}
with the mass $m_\alpha$, three momentum $\vec{q}$ of the incoming meson
$\alpha$ ($=\pi$, $\rho$, $\omega$), and the cutoff parameter $\Lambda$.
Here, we introduce two cutoff parameters for a heavy meson
($D^{(\ast)}$ and $\bar{B}^{(\ast)}$) and a nucleon.
The cutoff parameter $\Lambda_N$ for the nucleon is determined to
reproduce the properties (the binding energy, the scattering length and effective range) of the deuteron.
The cutoff parameter $\Lambda_P$ for heavy meson $P^{(\ast)}=D^{(\ast)}$, $\bar{B}^{(\ast)}$ is
determined by the ratios of matter radii of the heavy meson and nucleon
where the ratio is obtained by a quark model calculation~\cite{Yasui:2009bz,Yamaguchi:2011xb,Yamaguchi:2011qw}.
Their numerical values are summarized in Table~\ref{cutoff}.

In the present study,
we employ two types of potential: the $\pi$ exchange and
the $\pi\rho\,\omega$ exchanges.
By comparing the results from them, we show the importance of the one pion
exchange potential.

\begin{table}[htbp]
 \caption{Cutoff parameters of a nucleon ($\Lambda_N$) and heavy
 mesons ($\Lambda_D$ for $D^{(\ast)}$ meson and $\Lambda_B$ for
 $\bar{B}^{(\ast)}$ meson) as employed in Ref.~\cite{Yamaguchi:2011xb,Yamaguchi:2011qw}.}
 \label{cutoff}
\begin{tabular}{lccc}
 \hline
 Potential&$\Lambda_N$ [MeV] &$\Lambda_D$ [MeV] &$\Lambda_B$ [MeV] \\
 \hline
 $\pi$&830&1121&1070 \\
 $\pi\rho\,\omega$&846&1142&1091 \\
 \hline
\end{tabular}
\end{table}

\begin{table}[htbp]
\centering
\caption{\small Various coupled channels for a 
given quantum number $J^P$.  }
\label{table_qnumbers}
\vspace*{0.5cm}
{\small 
\begin{tabular}{ c  | c c c c}
\hline
$J^P$ &  \multicolumn{4}{c }{channels} \\
\hline
$1/2^-$ &$PN(^2S_{1/2})$&$P^\ast N(^2S_{1/2})$&$P^\ast N(^4D_{1/2})$& \\
$1/2^+$ &$PN(^2P_{1/2})$&$P^\ast N(^2P_{1/2})$&$P^\ast N(^4P_{1/2})$& \\
$3/2^-$ & $PN(^2D_{3/2})$&$P^\ast N(^4S_{3/2})$&$P^\ast
	      N(^4D_{3/2})$&$P^\ast N(^2D_{3/2})$ \\
$3/2^+$ & $PN(^2P_{3/2})$&$P^\ast N(^2P_{3/2})$&$P^\ast
	      N(^4P_{3/2})$&$P^\ast N(^4F_{3/2})$ \\
$5/2^-$&$PN(^2D_{5/2})$&$P^\ast N(^2D_{5/2})$&$P^\ast
	      N(^4D_{5/2})$&$P^\ast N(^4G_{5/2})$ \\
$5/2^+$&$PN(^2F_{5/2})$&$P^\ast N(^4P_{5/2})$&$P^\ast
	      N(^2F_{5/2})$&$P^\ast N(^4F_{5/2})$ \\
$7/2^-$&$PN(^2G_{7/2})$&$P^\ast N(^4D_{7/2})$&$P^\ast
	      N(^2G_{7/2})$&$P^\ast N(^4G_{7/2})$ \\
$7/2^+$&$PN(^2F_{7/2})$&$P^\ast N(^2F_{7/2})$&$P^\ast
	      N(^4F_{7/2})$&$P^\ast N(^4H_{7/2})$ \\
\hline
\end{tabular}
}
\end{table}

\section{Numerical Results}
\label{result}
We consider the states with $J^P=1/2^\pm$, $3/2^\pm$, $5/2^\pm$ and $7/2^\pm$
(total angular momentum $J$ and parity $P$) with isospin $I=0$ and $1$.
Various states with $J^P$ are expanded by coupled channels of 
$^{2S+1}L_{J}$ (spin $S$ and orbital angular momentum $L$)
as summarized in Table.~\ref{table_qnumbers}.
The meson exchange potentials among various channels and kinetic terms are
summarized in Appendix~\ref{appendix_a}.
Using the hamiltonian composed of the kinetic and potential terms,
we solve the coupled-channel Schr\"odinger equations for $PN$ and
$P^\ast N$ channels numerically.

\subsection{Isospin singlet ($I=0$)}

We present the results for the isosinglet state ($I=0$).
The energies of bound and resonant states are summarized in 
Table~\ref{tableDN} and also presented in Figs.~\ref{FigenergylevelI=0} and \ref{FigenergylevelBI=0}.
The results are shown for the $\pi$ and $\pi \rho \, \omega$ potentials, and
the corresponding energy levels
are connected by arrows in Figs.~\ref{FigenergylevelI=0} and \ref{FigenergylevelBI=0}.
The numerical values are measured from the $DN$ and $\bar{B}N$ thresholds,
respectively.  

First we discuss the states with $J^P=1/2^\pm$ and
$3/2^-$, and then those with $J^P=3/2^+$, $5/2^\pm$ and $7/2^\pm$.
In fact, it will turn out that the results in those two categories have different behaviors.

\subsubsection{$J^P=1/2^\pm$ and $3/2^-$}

\begin{figure}[htbp]
 \begin{center}
  \includegraphics[width=15cm,clip]{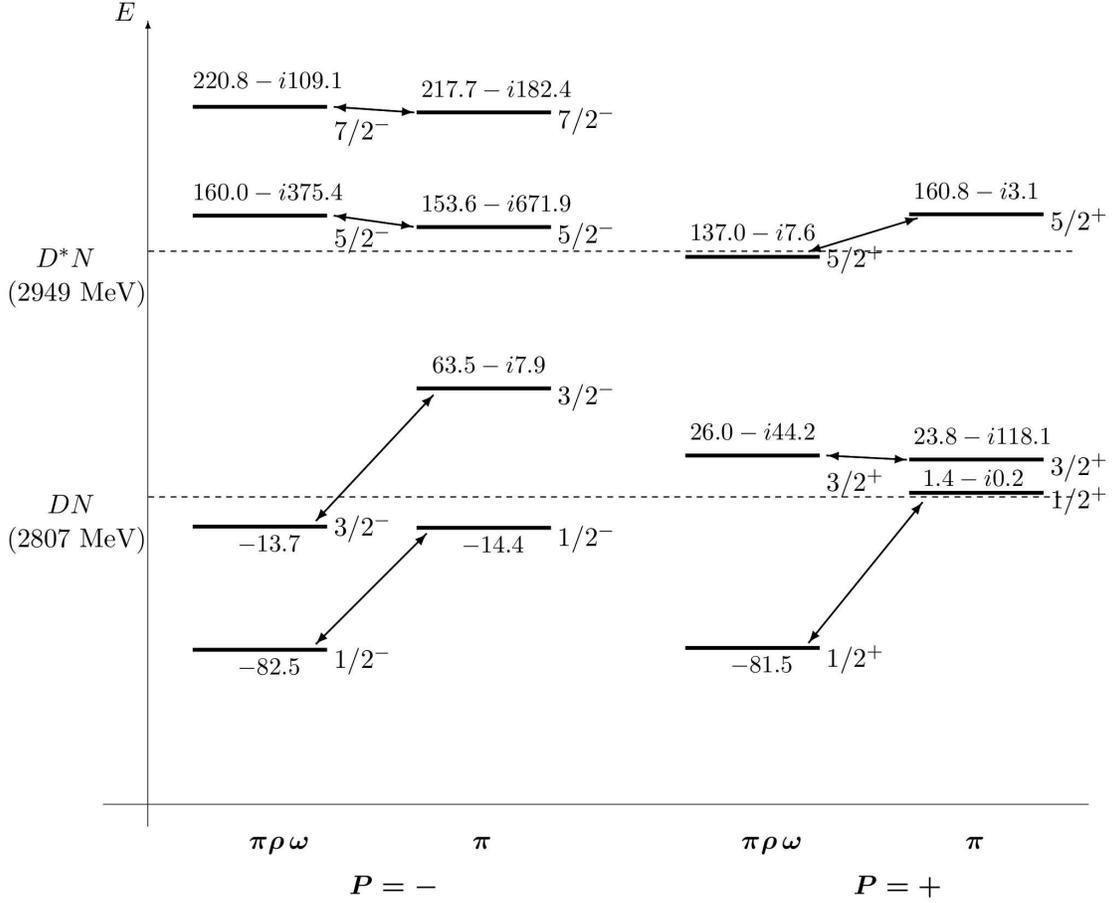}
 \end{center}
 \caption{Energies of bound and resonant states of $D^{(\ast)}N$ for $I=0$ with positive parity ($P=+$) and
 negative parity ($P=-$). The energies are measured from the $DN$
  threshold. The results for $\pi$ and $\pi\rho\,\omega$ potentials are
 shown, where corresponding states are connected by arrows.
 The binding energies are given as
  real negative value, and the resonance energies $E_{\mathrm{re}}$ and
  decay widths $\Gamma$ are given as $E_{\mathrm{re}}-i\Gamma/2$. 
 }
 \label{FigenergylevelI=0}
\end{figure}

\begin{figure}[htbp]
 \begin{center}
  \includegraphics[width=15cm,clip]{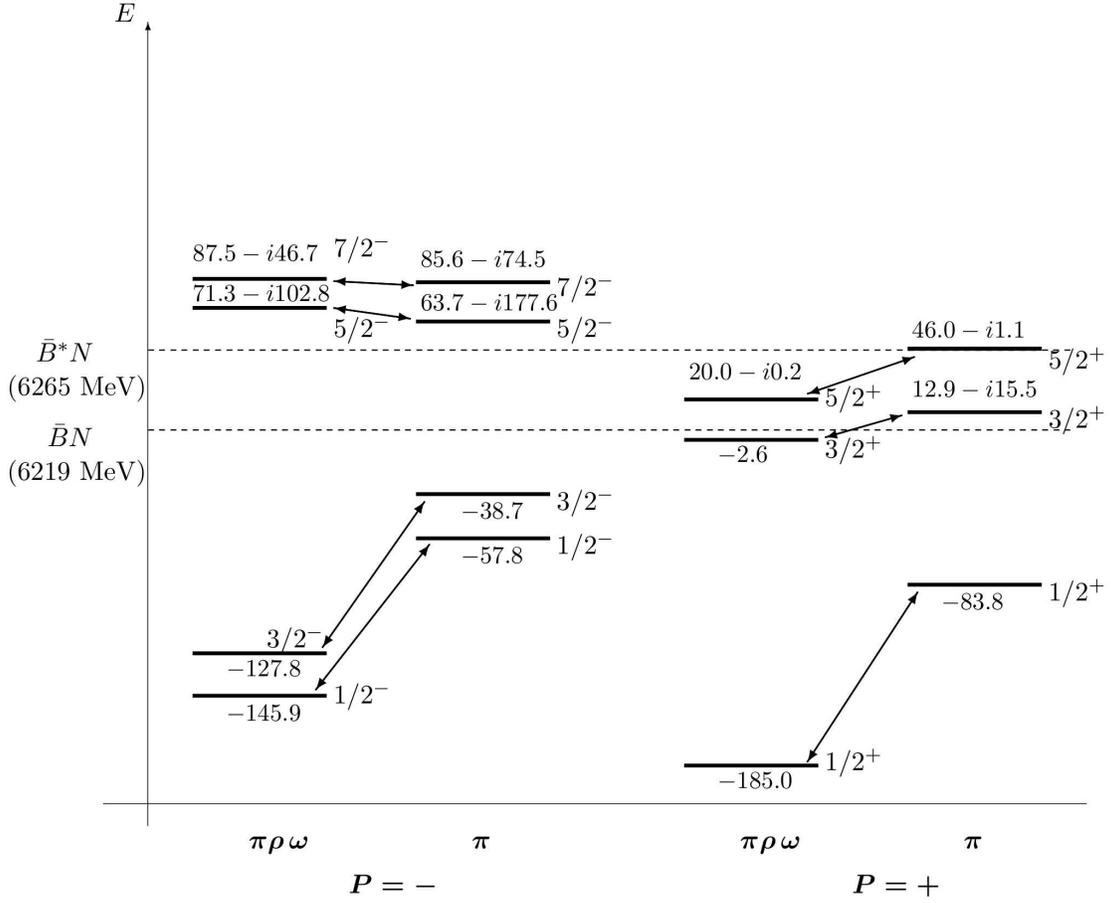}
 \end{center}
 \caption{Energies of bound and resonant states of $\bar{B}^{(\ast)}N$ for $I=0$ with positive parity ($P=+$) and
 negative parity ($P=-$). The energies are measured from the $\bar{B}N$
  threshold.
 The same convention is used as Fig~\ref{FigenergylevelI=0}.
 }
 \label{FigenergylevelBI=0}
\end{figure}

For $(I,J^P)=(0,1/2^-)$,
we find bound states in both of $D^{(\ast)}N$ and $\bar{B}^{(\ast)}N$.
For $D^{(\ast)}N$, the binding energies are $-14.4$ MeV for the $\pi$ potential and
$-82.5$ MeV for the $\pi\rho\,\omega$ potential.
The relative radii are 1.51 fm and 0.86 fm, respectively.
As expected, for a larger binding energy the system becomes smaller.  
The tensor force from the $\pi$ exchange which is the main driving force
of the $DN$-$D^{\ast}N$ mixing
plays an important role to form the bound states.
In fact, we have verified that  neither bound nor resonant state exists when the tensor force from the $\pi$ exchange is switched off and the $DN$-$D^\ast N$ mixing is small.
For $J^P=1/2^-$, the results in the $\pi$ and $\pi\rho\,\omega$
potentials are very different
as indicated in Figs.~\ref{FigenergylevelI=0} and \ref{FigenergylevelBI=0}.
Since both $\rho$ and $\omega$ exchanges are attractive for $D^{(\ast)}N$ ($\bar{B}^{(\ast)}N$)
system,
the vector meson exchanges contribute to form the deeply bound state
of the binding energy around 80 MeV.
This contrasts with the previous result for the $\bar{D}^{(\ast)}N$ and
$B^{(\ast)}N$ systems of truly exotic channels, where the $\rho$ and
$\omega$ exchanges play a minor role
due to the cancellation of these two potentials~\cite{Yamaguchi:2011xb,Yamaguchi:2011qw}.

We estimate the mixing ratio of various channels in the bound states as
summarized in Table~\ref{mixing}.
We observe that, for $J^P=1/2^-$ states, the most
dominant component is $DN(^2S_{1/2})$ with a fraction 71.8  \%.
The second dominant component is $D^{\ast}N(^{4}D_{1/2})$ with a fraction 20.4  \%.
Therefore, the tensor force which mixes the $S$-wave in $DN(^{2}S_{1/2})$ and the $D$-wave in $D^{\ast}N(^{4}D_{1/2})$ is important to provide a strong attraction.

For $\bar{B}^{(\ast)}N$ of $(I,J^P)=(0,1/2^-)$,
we also obtain bound states. 
The binding energies are $-57.8$ MeV for the $\pi$ potential and $-145.9$
MeV for the $\pi\rho\,\omega$ potential, and the relative radii are $0.99$ fm and
$0.76$ fm, respectively.
Again the results of the two potentials are very different with the same reason for the $D^{(\ast)}N$ system. 
As compared to the $D^{(\ast)}N$ system,
the binding energy in the $\bar{B}^{(\ast)}N$ system is  much larger,
because heavier particles suppress kinetic energy.
For the mixing ratios, we also find a similar tendency as we discussed for
the $D^{(\ast)}N$ system; 56.1  \% for $\bar{B}N(^2S_{1/2})$, 13.3  \% for
$\bar{B}^\ast N(^2S_{1/2})$ and 30.6  \% for $\bar{B}^\ast N(^4D_{1/2})$.

 \begin{table}[h]
  \caption{Properties of bound and resonant states of $D^{(\ast)}N$ and $\bar{B}^{(\ast)}N$ systems. The energies
  $E$ are either pure real for bound states or complex for
  resonant states. 
  The complex energies for
  resonances are written as $E_{\rm{re}}-i\Gamma/2$ where $E_{\rm{re}}$ is the
  resonance energy and $\Gamma/2$ is the half-width.
  The binding and resonance energies are measured from the lowest
  threshold ($DN$ and $\bar{B}N$).
  Root mean square radii are shown only for bound states. 
  }
  \label{tableDN}
  \vspace{-1mm}
  \begin{center}
 \begin{tabular}{c|c|cc||cc}
 \hline
  \multirow{2}{*}{$I(J^P)$}&\multirow{2}{*}{Potential}&\multicolumn{2}{c||}{$DN$}&\multicolumn{2}{c}{$\bar{B}N$}\\ \cline{3-6}
  &&$E$ [MeV] & $\langle{r^2}\rangle^{1/2}$ [fm]&$E$ [MeV] &$\langle{r^2}\rangle^{1/2}$ [fm]  \\\hline
  \multirow{2}{*}{$0(1/2^-)$}&$\pi$&$-14.4$& $1.51$&$-57.8$&$0.99$\\
  &$\pi\rho\,\omega$&$-82.5$&$0.86$&$-145.9$&$0.76$\\\hline
  \multirow{2}{*}{$0(1/2^+)$}&$\pi$&$1.4-i0.2$& --- &$-83.8$&$0.92$\\
  &$\pi\rho\,\omega$&$-81.5$&$0.85$&$-185.0$&$0.75$\\\hline
  \multirow{2}{*}{$0(3/2^-)$}&$\pi$&$63.5-i7.9$& --- &$-38.7$&$0.99$\\
  &$\pi\rho\,\omega$&$-13.7$&$0.89$&$-127.8$&$0.76$\\\hline
  \multirow{2}{*}{$0(3/2^+)$}&$\pi$&$23.8-i118.1$& --- &$12.9-i15.5$& --- \\
  &$\pi\rho\,\omega$&$26.0-i44.2$&--- &$-2.6$&$1.81$\\\hline
  \multirow{2}{*}{$0(5/2^-)$}&$\pi$&$153.6-i671.9$& --- &$63.7-i177.6$&---\\
  &$\pi\rho\,\omega$&$160.0-i375.4$&--- &$71.3-i102.8$&---\\\hline
  \multirow{2}{*}{$0(5/2^+)$}&$\pi$&$160.8-i3.1$& --- &$46.0-i1.1$&---\\
  &$\pi\rho\,\omega$&$137.0-i7.6$&--- &$20.0-i0.2$&---\\\hline
  \multirow{2}{*}{$0(7/2^-)$}&$\pi$&$217.7-i182.4$& --- &$85.6-i74.5$&---\\
  &$\pi\rho\,\omega$&$220.8-i109.1$ & --- &$87.5-i46.7$&\\\hline
  \multirow{2}{*}{$0(7/2^+)$}&$\pi$&no& --- &no&---\\
  &$\pi\rho\,\omega$&no& --- &no&---\\\hline\hline
  \multirow{2}{*}{$1(1/2^-)$}&$\pi$& no & --- &no&---\\
  &$\pi\rho\,\omega$&$147.2-i105.5$& ---&$50.7-i75.5$&---\\
  \hline
 \end{tabular}
  \end{center}
 \end{table}

\begin{table}[htbp]
 \caption{Mixing ratio of each channel in the bound $D^{(\ast)}N$ and
 $\bar{B}^{(\ast)}N$ states 
 for $J^P=1/2^\pm$ and $3/2^\pm$ with $I=0$ when the
 $\pi\rho\,\omega$ potential is employed.}
 \label{mixing}
 \begin{center}
  \begin{tabular}{c|cccc||c|cccc}
   \hline
   $1/2^-$ &\scalebox{0.95}{$PN(^2S_{1/2})$} &\scalebox{0.95}{$P^\ast N(^2S_{1/2})$} &\scalebox{0.95}{$P^\ast N(^4D_{1/2})$}&---&
   $1/2^+$ &\scalebox{0.95}{$PN(^2P_{1/2})$} &\scalebox{0.95}{$P^\ast N(^2P_{1/2})$} &\scalebox{0.95}{$P^\ast N(^4P_{1/2})$}&--- \\
   \hline
   $DN$&$71.8 \%$ &$7.8 \%$ &$20.4 \%$&---&
   $DN$&$38.8 \%$ &$6.0 \%$ &$55.2 \%$&--- \\
   $\bar{B}N$&$56.1 \%$ &$13.3 \%$ &$30.6 \%$&---&
   $\bar{B}N$&$28.4 \%$ &$7.7 \%$ &$63.9 \%$&--- \\
   \hline\hline
   $3/2^-$ &\scalebox{0.95}{$PN(^2D_{3/2})$} &\scalebox{0.95}{$P^\ast N(^4S_{3/2})$} &\scalebox{0.95}{$P^\ast N(^4D_{3/2})$} &\scalebox{0.95}{$P^\ast N(^2D_{3/2})$}&
   $3/2^+$ &\scalebox{0.95}{$PN(^2P_{3/2})$} &\scalebox{0.95}{$P^\ast N(^2P_{3/2})$} &\scalebox{0.95}{$P^\ast N(^4P_{3/2})$} &\scalebox{0.95}{$P^\ast N(^4F_{3/2})$} \\
   \hline
   $DN$&$19.8 \%$ &$62.8 \%$ &$14.2 \%$&$3.2 \%$&
   $DN$&--- &--- &---&--- \\
   $\bar{B}N$&$14.6 \%$ &$64.7 \%$ &$16.7 \%$&$4.0 \%$&
   $\bar{B}N$&$71.2 \%$ &$6.7 \%$ &$7.5 \%$&$14.6 \%$ \\
   \hline
  \end{tabular}
 \end{center}
\end{table}


Let us move to $(I,J^P)=(0,1/2^+)$ state.
For $D^{(\ast)}N$, we find one resonance near the $DN$ threshold when the $\pi$ potential is used.
The resonance energy is $1.4$ MeV and the half decay width is $0.2$ MeV as shown in Table~\ref{tableDN}.
In the scattering state, we define the resonance energy $E_{\rm{re}}$ by an inflection point of
the phase shift~\cite{Arai:1999pg}, and 
the decay width $\Gamma$ is given by $\Gamma=2/(d\delta/dE)_{E=E_{\rm{re}}}$.
When the $\pi\rho\,\omega$ potential is used, 
we find a bound state with a binding
energy $-81.5$ MeV and with a relative radius $0.85$ fm.
The mixing ratios of the bound state are 38.8 \% for $DN(^2P_{1/2})$, 6.0 \% for 
$D^\ast N(^2P_{1/2})$ and 55.2 \% for $D^\ast N(^4P_{1/2})$, as shown in Table.~\ref{mixing}.
Interestingly, 
$D^\ast N(^4P_{1/2})$ is the most dominant channel, although the mass of $D^\ast N$ is heavier than the mass of $DN$.
This is because the attraction of tensor force is the strongest for $D^\ast N(^4P_{1/2})$ channel.

For $\bar{B}^{(\ast)}N$ of $(I,J^P)=(0,1/2^+)$,
we find bound states for both cases when the $\pi$ and $\pi\rho\,\omega$ potentials are used.
The binding energies are $-83.8$ MeV for the $\pi$
potential and $-185.0$ MeV for the $\pi\rho\,\omega$ potential, and the
relative radii are $0.92$ fm and $0.75$ fm, respectively.
For the $\pi\rho\,\omega$ potential, the mixing ratios of the bound
state are 28.4 \% for $\bar{B}N(^2P_{1/2})$, 7.7 \% for
$\bar{B}^\ast N(^2P_{1/2})$ and 63.9 \% for $\bar{B}^\ast N(^4P_{1/2})$.
In this case again, $\bar{B}^\ast N(^4P_{1/2})$ is the most dominant component regardless of its heavy mass.

For $(I,J^P)=(0,3/2^-)$,
we find a resonant $DN$ state for the $\pi$ potential with the resonance energy $63.5$ MeV and
the half decay width $7.9$ MeV as shown in Table~\ref{tableDN}.
In contrast, for the $\pi\rho\,\omega$ potential, we find a bound $DN$
state with the binding energy $-13.7$ MeV and the relative radius $0.89$
fm.
For the bound state, the mixing ratios are 19.8 \% for $DN(^2D_{3/2})$, 62.8 \% for $D^\ast N(^4S_{3/2})$,
14.2 \% for $D^\ast N(^4D_{3/2})$ and 3.2 \% for $D^\ast N(^2D_{3/2})$.
Thus, $D^\ast N(^4S_{3/2})$ is the dominant channel although its mass is heavier than the mass of $DN(^2D_{3/2})$;
once again a large attraction due to the tensor force is provided.

For $\bar{B}^{(\ast)}N$ state of $(I,J^P)=(0,3/2^-)$,
we obtain bound states for both cases the $\pi$ and
$\pi\rho\,\omega$ potentials are used.
The binding energies are $-38.7$ MeV for the $\pi$
potential and $-127.8$ MeV for the $\pi\rho\,\omega$ potential
with relative radii $0.99$ fm and $0.76$ fm, respectively.

\subsubsection{$J^P=3/2^+$, $5/2^{\pm}$ and $7/2^{\pm}$}

For $(I,J^P)=(0,3/2^+)$,
we find resonant states above the threshold.
For $D^{(\ast)}N$, the resonance energies are $23.8$ MeV for the $\pi$ potential and $26.0$ MeV 
for the $\pi\rho\,\omega$ potential.
The half decay widths are $118.1$ MeV and $44.2$ MeV, respectively.
As compared to the cases of $1/2^\pm$ and $3/2^-$,
the results of the $\pi$ and $\pi\rho\,\omega$ potentials are not very
much different, as indicated in Fig.~\ref{FigenergylevelI=0}. 
Since the wave functions 
are extended due to larger orbital angular momenta of $P$-wave ($L=1$) and $F$-wave ($L=3$) for $J^P=3/2^+$, 
the long range potential of the $\pi$ exchange dominates, while 
the short range potentials from $\rho$ and $\omega$ exchanges are suppressed.
For $\bar{B}^{(\ast)}N$,
when the $\pi$ potential is used, we find a resonant state whose resonance energy is
$12.9$ MeV and half decay width is $15.5$ MeV, while
when the $\pi\rho\,\omega$ potential is used, we find a loosely bound
state
of a binding energy $-2.6$ MeV and a relative radius $1.81$ fm.
The mixing ratios are 71.2 \% for $\bar{B}N(^2P_{3/2})$, 6.7 \% for
$\bar{B}^\ast N(^2P_{3/2})$, 7.5 \% for $\bar{B}^\ast N(^4P_{3/2})$ and
14.6 \% for $\bar{B}^\ast N(^4F_{3/2})$.
The most dominant component is $\bar{B}N(^2P_{3/2})$, 
and the second dominant one is $\bar{B}^\ast N(^4F_{3/2})$. 

For $(I,J^P)=(0,5/2^-)$,
we obtain resonant states for both cases when the $\pi$ and $\pi\rho\,\omega$
potentials are used. 
For $D^{(\ast)}N$, the resonance energies are 153.6 MeV for the $\pi$ potential and
160.0 MeV for the $\pi\rho\,\omega$ potential,
which are above the $D^\ast N$ threshold.
The corresponding half decay widths are 671.9 MeV and 375.4 MeV, respectively.
The difference between the results of the $\pi$ and
$\pi\rho\,\omega$ potentials is once again small,
due to the same reason as before with large angular momenta.
For $\bar{B}^{(\ast)}N$ state, we also find resonant states above the $\bar{B}^\ast N$
threshold.
The resonance positions are $63.7$ MeV for the $\pi$ potential and $71.3$ MeV for the
$\pi\rho\,\omega$ potential.
The corresponding half decay widths are $177.6$ MeV and $102.8$ MeV, respectively.

For $(I,J^P)=(0,5/2^+)$,
we find resonant states with narrow widths for both cases when the $\pi$ and $\pi\rho\,\omega$ potentials are used.
For $D^{(\ast)}N$,
when the $\pi$ potential is used, 
the resonance energy is $160.8$ MeV and the half decay width $\Gamma/2=3.1$ MeV.
When the $\pi\rho\,\omega$ potential is used,
the resonance energy is $137.0$ MeV and 
the half decay width $\Gamma/2=7.6$ MeV.
For $\bar{B}^{(\ast)}N$,
we also find resonances whose energies are $46.0$ MeV
for the $\pi$
potential and $20.0$ MeV for the $\pi\rho\,\omega$ potential,
with the corresponding half decay widths $1.1$ MeV and $0.2$ MeV, respectively.
Again, the results for the $\pi$ and $\pi\rho\,\omega$ potentials are similar.

For $(I,J^P)=(0,7/2^-)$,
we obtain resonances above the $D^\ast N$ and $\bar{B}^\ast N$ thresholds.
For $D^{(\ast)} N$,
there exist resonances at 217.7 MeV for the $\pi$ potential and
at 220.8 MeV for the $\pi\rho\,\omega$ potential,
with half decay widths 182.4 MeV and 109.1 MeV, respectively.
For $\bar{B}^{(\ast)}N$,
there also exist resonances whose energies are 85.6 MeV
for the $\pi$ potential and 87.5 MeV
for the $\pi\rho\,\omega$ potential,
with half decay widths $74.5$ MeV and $46.7$ MeV, respectively.

Finally, for $(I,J^P)=(0,7/2^+)$, we find no bound nor resonant state.

\begin{figure}[htbp]
 \begin{center}
  \includegraphics[width=15cm,clip]{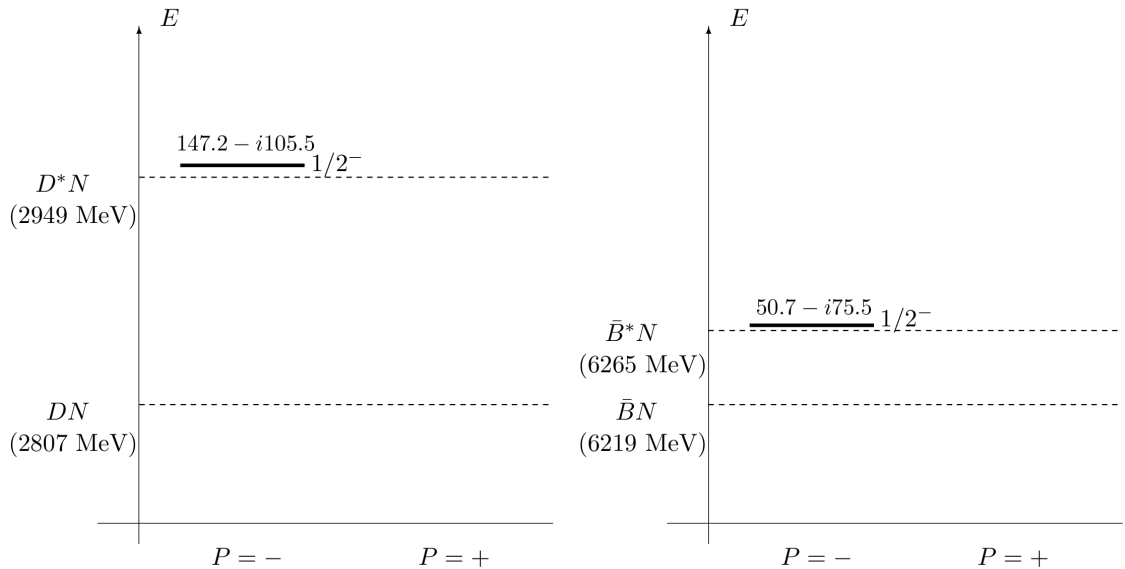}
 \end{center}
 \caption{Energies of bound and resonant states of $D^{(\ast)}N$ and
 $\bar{B}^{(\ast)}N$ for $I=1$ when the $\pi\rho\,\omega$ potential is used. The same convention is used as Fig.\ref{FigenergylevelI=0}.
\label{FigenergylevelI=1}
}
\end{figure}

\subsection{Isospin triplet ($I=1$)}

Let us move to the states of isospin triplet ($I=1$).
We summarize the results for $D^{(\ast)}N$ and for $\bar{B}^{(\ast)}N$ in Table~\ref{tableDN}
and show the energy levels in Fig.~\ref{FigenergylevelI=1}.
As a result, we find resonant states only for $J^P=1/2^-$ when
the $\pi\rho\,\omega$ potential is employed.
For $D^{(\ast)}N$, the resonance energy is $147.2$ MeV and the half decay width is
$105.5$ MeV.
For $\bar{B}^{(\ast)}N$ state, we obtain the resonance whose energy is $50.7$
MeV and half width $75.5$ MeV.
The reason that there are not many states in isospin triplet channel can be
understood as follows;
as compared to the isosinglet channel,
the attractive force in the isotriplet channel
is weak due to small isospin factor;
$\vec{\tau}_P\cdot\vec{\tau}_N=-3$ for isosinglet and
$\vec{\tau}_P\cdot\vec{\tau}_N=1$ for isotriplet.

\section{Discussion}
\label{Discussion}

\subsection{Contribution from short range interactions}
By solving coupled channel Schr\"odinger equations for $PN$ and 
$P^\ast N$ channels, we find many bound and resonant states for $I=0$ and few
resonances for $I=1$.
For these states,
the tensor force of the $\pi$ exchange
potential plays a significant role to produce them.
When we ignore the $P^\ast N$ channels and solve the Schr\"odinger equation only
with $PN$ channels,
we find no bound state nor resonance.
Thus, the $PN$-$P^\ast N$ mixing and the $\pi$ exchange potential play an important
role to generate a rich structure of molecular states.

The importance of the $\pi$ exchange potential stands out
more in the bottom sector.
The small mass difference between $\bar{B}$ and $\bar{B}^\ast$ mesons helps to yield
the strong attraction because it induces the strong $\bar{B}N$-$\bar{B}^\ast N$ mixing
with the tensor force.
Furthermore, the $\bar{B}^{(\ast)}N$ has heavier reduced mass and
hence the kinetic term is suppressed.
For these reasons, the binding energies of $\bar{B}^{(\ast)}N$ states are
larger than these of $D^{(\ast)}N$ states.

When we compare the results of the $\pi$ potential and of the
$\pi\rho\,\omega$ potential,
they are quite different for $J^P=1/2^\pm$, $3/2^-$ with $I=0$, where
the $\rho$ and $\omega$ exchange potentials become important to produce
attraction.
For $D^{(\ast)}N$ and $\bar{B}^{(\ast)}N$ states,
both $\rho$ and $\omega$ exchange potentials are attractive,
and hence they increase the binding energy significantly.
On the other hand, for $J^P=3/2^+$, $5/2^\pm$ and $7/2^-$,
the results for the $\pi\rho\,\omega$ potential are
similar to those for the $\pi$ potential.
For large $J$ states, the $\pi$ exchange potential plays a dominant role to
generate bound and resonant states,
while the $\rho$ and $\omega$ exchanges play only a minor one.
This is attributed to the fact that these states 
have large orbital angular momenta. 
If relevant channels include large orbital angular momenta, 
the wave functions tend to extend spatially,
and the long range force, namely  the $\pi$ exchange
potential, becomes important while the short range force is suppressed.

\subsection{Emergent hadronic molecules}

In the present study, 
we discuss the molecular structure formed by the $P^{(\ast)}N$ bound and
resonant states.
However, the hadronic molecular picture is not applicable to deeply bound states with small
radii.
For such states,
the constituent hadrons, namely $D^{(\ast)}$ ($\bar{B}^{(\ast)}$) meson and nucleon,
overlap each other,
and therefore we have to consider short range effects such as an internal
structure of hadrons, channel couplings to conventional three quark
states and so on.
As a naive criterion for the hadronic molecule,
we have shown the relative radii of the bound states as discussed in Ref~\cite{Ohkoda:2012hv}.
If the size of the bound state is larger than twice of typical hadron size
(namely 1 fm),
the state could be well described by a molecular structure.
For resonant states, we identify the states as the hadronic molecule.
According to the criterion, for the $\pi\rho\,\omega$ potential,
only the bound state for $J^P=3/2^+$ of $\bar{B}^{(\ast)}N$ constructs a hadronic
molecule, where the relative radius is 1.81 fm.
Contrary to this,
the bound states for $J^P=1/2^\pm$ and $3/2^-$ with $I=0$ which have a small
radius and a large binding energy are not described as simple molecules.
We need to consider
the short range effects 
including various channel couplings to do more realistic discussions.

\section{Summary}

We have investigated heavy baryons as hadronic molecules formed by a
heavy meson and a nucleon, $D^{(\ast)}N$ and $\bar{B}^{(\ast)}N$.
The interaction is given by the meson exchange potential with respecting
the heavy quark and chiral symmetries.
Many bound states and resonances are found in isosinglet channel both
for $D^{(\ast)}N$ and $\bar{B}^{(\ast)}N$ systems.
In contrast, there are few resonances in isotriplet channel due to small isospin
factor.
For $J^P=1/2^\pm$ and $3/2^-$,
we have found deeply bound states far below
$DN$ or $\bar{B}N$ threshold with a small radius.
In these states, the vector meson exchange potential
yields a strong attraction.
In order to perform more realistic discussions for such compact states, we
need to consider further effects
from internal structures of constituent hadrons and channel couplings to quark intrinsic states.
On the other hand, for $J^P=3/2^+$, $5/2^\pm$ and $7/2^-$ with
large orbital angular momenta,
the one pion exchange potential is favored and we have obtained 
loosely bound states and resonances near the thresholds.
There the short range interactions are rather inactive and the long
range force of the one pion exchange dominates, where the tensor force
plays an important role to generate
these states, and they compose the hadronic molecular structure.

It is expected that the hadronic molecular states from a heavy meson and
a nucleon are studied in many accelerator facilities; 
the $D$ meson production in the antiproton-nucleon annihilation process
in FAIR~\cite{Wiedner:2011mf} and the heavy ion collision in RHIC and LHC~\cite{Cho:2010db,Cho:2011ew}.
Furthermore, the exotic baryons are investigated in J-PARC in the coming future. 

\section*{Acknowledgments}
This work is supported in part by Grant-in-Aid for Scientific Research
on Priority Areas ``Elucidation of New Hadrons with a Variety of
Flavors(E01: 21105006)''(S.~Y. and A.~H.) from the Ministry of Education, Culture,
Sports, Science and Technology of Japan and Grant-in-Aid for ``JSPS
Fellows(24-3518)''(Y.~Y.) from Japan Society for the Promotion of Science.
\appendix
\section{Potentials and kinetic terms}
\label{appendix_a}
Interaction potentials are derived by using the Lagrangians
in Eqs.~\eqref{LpiHH}, \eqref{LvHH}, \eqref{LpiNN} and \eqref{LvNN} as
shown in Ref.~\cite{Yamaguchi:2011xb,Yamaguchi:2011qw}.
The potentials for the coupled channel systems are given in the matrix
form of $3\times 3$ for $J^P=1/2^\pm$ and of $4\times 4$ for the other states.

The $\pi$ exchange potentials for each $J^P$ are obtained by
\begin{align}
 V^{\pi}_{1/2^-}&=
\left(
\begin{array}{ccc}
  0 & \sqrt{3} V^{\pi}_C & -\sqrt{6}V^{\pi}_T  \\
\sqrt{3}V^{\pi}_C & -2V^{\pi}_C & -\sqrt{2} V^{\pi}_T \\
-\sqrt{6}V^{\pi}_T & -\sqrt{2}V^{\pi}_T & V^{\pi}_C - 2V^{\pi}_T
\end{array}
\right) ,
\label{matpi1/2-}
\end{align}

\begin{align}
 V^{\pi}_{1/2^+}&=
\left(
\begin{array}{ccc}
  0 & \sqrt{3} V^{\pi}_C & -\sqrt{6}V^{\pi}_T  \\
\sqrt{3}V^{\pi}_C & -2V^{\pi}_C & -\sqrt{2} V^{\pi}_T \\
-\sqrt{6}V^{\pi}_T & -\sqrt{2}V^{\pi}_T & V^{\pi}_C - 2V^{\pi}_T
\end{array}
\right) ,
\label{matpi1/2+}
\end{align}

\begin{align}
  V^{\pi}_{3/2^-}&=
\left(
\begin{array}{cccc}
 0& \sqrt{3}V^{\pi}_T & -\sqrt{3}V^{\pi}_T& \sqrt{3}V^{\pi}_C \\
 \sqrt{3}V^{\pi}_T & V^{\pi}_C & 2V^{\pi}_T & V^{\pi}_T \\
-\sqrt{3}V^{\pi}_T & 2V^{\pi}_T & V^{\pi}_C & -V^{\pi}_T \\
 \sqrt{3}V^{\pi}_C & V^{\pi}_T & -V^{\pi}_T & -2V^{\pi}_C \\
\end{array}
\right) ,
\label{matpi3/2-}
\end{align}

\begin{align}
  V^{\pi}_{3/2^+}&=
\left(
\begin{array}{cccc}
 0& \sqrt{3}V^{\pi}_C &\displaystyle \sqrt{\frac{3}{5}}V^{\pi}_T&\displaystyle -3\sqrt{\frac{3}{5}}V^{\pi}_T \\
 \sqrt{3}V^{\pi}_C & -2V^{\pi}_C &\displaystyle {\frac{1}{\sqrt{5}}}V^{\pi}_T & -\sqrt{3}V^{\pi}_T \\
\displaystyle \sqrt{\frac{3}{5}}V^{\pi}_T &\displaystyle {\frac{1}{\sqrt{5}}}V^{\pi}_T &\displaystyle V^{\pi}_C+\frac{8}{5}V^{\pi}_T &\displaystyle \frac{6}{5}V^{\pi}_T \\
\displaystyle -3\sqrt{\frac{3}{5}}V^{\pi}_T & -\sqrt{3}V^{\pi}_T &\displaystyle \frac{6}{5}V^{\pi}_T &\displaystyle V^{\pi}_C-\frac{8}{5}V^{\pi}_T \\
\end{array}
\right) ,
\label{matpi3/2+}
\end{align}

\begin{align}
   V^{\pi}_{5/2^-}&=
\left(
\begin{array}{cccc}
 0& \sqrt{3}V^{\pi}_C&\displaystyle  \sqrt{\frac{6}{7}}V^{\pi}_T&\displaystyle  -\frac{6}{\sqrt{7}}V^{\pi}_T \\
 \sqrt{3}V^{\pi}_C& -2V^{\pi}_C&\displaystyle \sqrt{\frac{2}{7}}V^{\pi}_T&\displaystyle -2\sqrt{\frac{3}{7}}V^{\pi}_T \\
\displaystyle \sqrt{\frac{6}{7}}V^{\pi}_T&\displaystyle \sqrt{\frac{2}{7}}V^{\pi}_T&\displaystyle V^{\pi}_C+\frac{10}{7}V^{\pi}_T&\displaystyle \frac{4}{7}\sqrt{6}V^{\pi}_T\\
\displaystyle -\frac{6}{\sqrt{7}}V^{\pi}_T&\displaystyle -2\sqrt{\frac{3}{7}}V^{\pi}_T&\displaystyle \frac{4}{7}\sqrt{6}V^{\pi}_T&\displaystyle V^{\pi}_C-\frac{10}{7}V^{\pi}_T\\
\end{array}
\right) ,
\label{matpi5/2-}
\end{align}

\begin{align}
   V^{\pi}_{5/2^+}&=
\left(
\begin{array}{cccc}
 0&\displaystyle \frac{3}{5}\sqrt{10}V^{\pi}_T& \sqrt{3}V^{\pi}_C&\displaystyle -2\sqrt{\frac{3}{5}}V^{\pi}_T\\
\displaystyle \frac{3}{5}\sqrt{10}V^{\pi}_T&\displaystyle
 V^{\pi}_C-\frac{2}{5}V^{\pi}_T&\displaystyle
 \sqrt{\frac{6}{5}}V^{\pi}_T&\displaystyle \frac{4}{5}\sqrt{6}V^{\pi}_T\\
 \sqrt{3}V^{\pi}_C&\displaystyle
  \sqrt{\frac{6}{5}}V^{\pi}_T&-2V^{\pi}_C&\displaystyle -\frac{2}{\sqrt{5}}V^{\pi}_T\\
\displaystyle -2\sqrt{\frac{3}{5}}V^{\pi}_T&\displaystyle
 \frac{4}{5}\sqrt{6}V^{\pi}_T&\displaystyle
 -\frac{2}{\sqrt{5}}V^{\pi}_T&\displaystyle V^{\pi}_C+\frac{2}{5}V^{\pi}_T\\
\end{array}
\right) ,
\label{matpi5/2+}
\end{align}

\begin{align}
 V^{\pi}_{7/2^-}&=
\left(
\begin{array}{cccc}
 0&\displaystyle 3\sqrt{\frac{3}{7}}V^{\pi}_T& \sqrt{3}V^{\pi}_C&\displaystyle -\sqrt{\frac{15}{7}}V^{\pi}_T \\
\displaystyle 3\sqrt{\frac{3}{7}}V^{\pi}_T&\displaystyle V^{\pi}_C-\frac{4}{7}V^{\pi}_T&
\displaystyle  \frac{3}{\sqrt{7}}V^{\pi}_T&\displaystyle \frac{6}{7}\sqrt{5}V^{\pi}_T\\
 \sqrt{3}V^{\pi}_C&\displaystyle \frac{3}{\sqrt{7}}V^{\pi}_T& -2V^{\pi}_C&\displaystyle -\sqrt{\frac{5}{7}}V^{\pi}_T\\
\displaystyle -\sqrt{\frac{15}{7}}V^{\pi}_T&\displaystyle
 \frac{6}{7}\sqrt{5}V^{\pi}_T&\displaystyle
 -\sqrt{\frac{5}{7}}V^{\pi}_T&\displaystyle V^{\pi}_C+\frac{4}{7}V^{\pi}_T\\
\end{array}
\right) ,
\label{matpi7/2-}
\end{align}

\begin{align}
 V^{\pi}_{7/2^+}&=
\left(
\begin{array}{cccc}
 0& \sqrt{3}V^{\pi}_C& V^{\pi}_T& -\sqrt{5}V^{\pi}_T \\
 \sqrt{3}V^{\pi}_C& -2V^{\pi}_C&\displaystyle \frac{1}{\sqrt{3}}V^{\pi}_T&\displaystyle -\sqrt{\frac{5}{3}}V^{\pi}_T\\
 V^{\pi}_T&\displaystyle \frac{1}{\sqrt{3}}V^{\pi}_T&\displaystyle V^{\pi}_C+\frac{4}{3}V^{\pi}_T&\displaystyle \frac{2}{3}\sqrt{5}V^{\pi}_T\\
 -\sqrt{5}V^{\pi}_T&\displaystyle -\sqrt{\frac{5}{3}}V^{\pi}_T&
\displaystyle  \frac{2}{3}\sqrt{5}V^{\pi}_T&\displaystyle V^{\pi}_C-\frac{4}{3}V^{\pi}_T\\
\end{array}
\right) ,
\label{matpi7/2+}
\end{align}
where 
\begin{align}
 V^{\pi}_C&
=\frac{g_{\pi} g_{\pi NN}}{\sqrt{2}m_N f_{\pi}}\frac{1}{3}C_{m_\pi}\vec{\tau}_{P} \cdot \vec{\tau}_N ,\quad
  V^{\pi}_T
=\frac{g_{\pi} g_{\pi NN}}{\sqrt{2}m_N f_{\pi}}\frac{1}{3}T_{m_\pi}\vec{\tau}_{P} \cdot \vec{\tau}_N .
\end{align}
The $\vec{\tau}_P$ and $\vec{\tau}_N$
are the isospin matrices for $P^{(\ast)}$ and $N$.
The functions $C_m=C(r;m)$ and $T_m=T(r;m)$ are given by
\begin{align}
 C(r;m) & = \int \frac{d^3 q}{(2 \pi)^3} \frac{m^2}{\vec{q }\,^2 + m^2}
 e^{i\vec{q}\cdot \vec{r}} F(\Lambda_P,\vec{q}\,)F(\Lambda_N,\vec{q}\,) ,    \label{C} \\ 
T(r;m) S_{12}(\hat{r}) & = \int  \frac{d^3 q}{(2 \pi)^3} 
\frac{-\vec{q}\,^2}{\vec{q}\,^2 + m^2} S_{12}(\hat{q})
 e^{i\vec{q}\cdot \vec{r}} F(\Lambda_P,\vec{q}\,)F(\Lambda_N,\vec{q}\,) ,
\label{T}
\end{align}
with $S_{12}(\hat{x}) = 3(\vec{\sigma}_1 \cdot \hat{x}) (\vec{\sigma}_2
\cdot \hat{x}) -\vec{\sigma}_1 \cdot \vec{\sigma}_2$, and 
$F(\Lambda,\vec{q}\,)$ denotes the form factor given 
in Eq.~\eqref{formfactor}.

The vector meson exchange potentials ($v=\rho$, $\omega$) are
\begin{align}
  V^{v}_{1/2^-} = &
 \begin{pmatrix}
  V^{v\,\prime}_C & 2\sqrt{3} V^{v}_C & \sqrt{6}V^{v}_T  \\
  2\sqrt{3}V^{v}_C & V^{v\,\prime}_C-4V^{v}_C & \sqrt{2} V^{v}_T \\
  \sqrt{6}V^{v}_T & \sqrt{2}V^{v}_T & V^{v\,\prime}_C+2V^{v}_C + 2V^{v}_T
 \end{pmatrix}
,
\label{matrho1/2-}
\end{align}

\begin{align}
 V^{v}_{1/2^+} = & 
 \begin{pmatrix}
  V^{v\,\prime}_C & 2\sqrt{3} V^{v}_C & \sqrt{6}V^{v}_T  \\
  2\sqrt{3}V^{v}_C & V^{v\,\prime}_C-4V^{v}_C & \sqrt{2} V^{v}_T \\
  \sqrt{6}V^{v}_T & \sqrt{2}V^{v}_T & V^{v\,\prime}_C+2V^{v}_C + 2V^{v}_T
 \end{pmatrix}
,
\label{matrho1/2+}
\end{align}
\begin{align}
 V^{v}_{3/2^-} = &
\begin{pmatrix}
 V^{v\,\prime}_C & -\sqrt{3}V^{v}_T & \sqrt{3}V^{v}_T &2\sqrt{3}V^{v}_C \\
-\sqrt{3}V^{v}_T & V^{v\,\prime}_C+2V^{v}_C & -2V^{v}_T & -V^{v}_T \\
 \sqrt{3}V^{v}_T & -2V^{v}_T & V^{v\,\prime}_C+2V^{v}_C & V^{v}_T \\
 2\sqrt{3}V^{v}_C & -V^{v}_T & V^{v}_T & V^{v\,\prime}_C-4V^{v}_C \\
\end{pmatrix}
 ,
\label{matrho3/2-}
\end{align}
\begin{align}
 V^{v}_{3/2^+} = &
\begin{pmatrix}
 V^{v\,\prime}_C &  2\sqrt{3}V^{v}_C &\displaystyle -\sqrt{\frac{3}{5}}V^{v}_T &\displaystyle 3\sqrt{\frac{3}{5}}V^{v}_T \\
 2\sqrt{3}V^{v}_C & V^{v\,\prime}_C-4V^{v}_C &\displaystyle -\frac{1}{\sqrt{5}}V^{v}_T &\displaystyle \frac{3}{\sqrt{5}}V^{v}_T \\
\displaystyle -\sqrt{\frac{3}{5}}V^{v}_T &\displaystyle
 -\frac{1}{\sqrt{5}}V^{v}_T &\displaystyle  V^{v\,\prime}_C+2V^{v}_C-\frac{8}{5}V^{v}_T &\displaystyle -\frac{6}{5}V^{v}_T \\
\displaystyle 3\sqrt{\frac{3}{5}}V^{v}_T &\displaystyle \frac{3}{\sqrt{5}}V^{v}_T &\displaystyle -\frac{6}{5}V^{v}_T &\displaystyle V^{v\,\prime}_C+2V^{v}_C+\frac{8}{5}V^{v}_T
\end{pmatrix}
,
\label{matrho3/2+}
\end{align}

\begin{align}
 V^{v}_{5/2^-} = &
\begin{pmatrix}
 V^{v\,\prime}_C & 2\sqrt{3}V^{v}_C&\displaystyle
 -\sqrt{\frac{6}{7}}V^{v}_T&\displaystyle \frac{6}{\sqrt{7}}V^{v}_T\\
 2\sqrt{3}V^{v}_C&V^{v\,\prime}_C-4V^{v}_C&\displaystyle -\sqrt{\frac{2}{7}}V^{v}_T&\displaystyle 2\sqrt{\frac{3}{7}}V^{v}_T\\
\displaystyle  -\sqrt{\frac{6}{7}}V^{v}_T&\displaystyle  -\sqrt{\frac{2}{7}}V^{v}_T&\displaystyle V^{v\,\prime}_C+2V^{v}_C-\frac{10}{7}V^{v}_T&\displaystyle -\frac{4}{7}\sqrt{6}V^{v}_T\\
\displaystyle  \frac{6}{\sqrt{7}}V^{v}_T&\displaystyle 2\sqrt{\frac{3}{7}}V^{v}_T&\displaystyle -\frac{4}{7}\sqrt{6}V^{v}_T&\displaystyle V^{v\,\prime}_C+2V^{v}_C+\frac{10}{7}V^{v}_T\\
\end{pmatrix}
,
\label{matrho5/2-}
\end{align}

\begin{align}
  V^{v}_{5/2^+} = &
\begin{pmatrix}
V^{v\,\prime}_C &\displaystyle -\frac{3}{5}\sqrt{10}V^{v}_T &2\sqrt{3}V^{v}_C
 &\displaystyle 2\sqrt{\frac{3}{5}}V^{v}_T \\
\displaystyle
 -\frac{3}{5}\sqrt{10}V^{v}_T&\displaystyle V^{v\,\prime}_C+2V^{v}_C+\frac{2}{5}V^{v}_T
 &\displaystyle -\sqrt{\frac{6}{5}}V^{v}_T &\displaystyle
 -\frac{4}{5}\sqrt{6}V^{v}_T \\
2\sqrt{3}V^{v}_C &\displaystyle
 -\sqrt{\frac{6}{5}}V^{v}_T&V^{v\,\prime}_C-4V^{v}_C &\displaystyle
 \frac{2}{\sqrt{5}}V^{v}_T \\
\displaystyle 2\sqrt{\frac{3}{5}}V^{v}_T &\displaystyle
 -\frac{4}{5}\sqrt{6}V^{v}_T &\displaystyle
 \frac{2}{\sqrt{5}}V^{v}_T&\displaystyle V^{v\,\prime}_C+2V^{v}_C-\frac{2}{5}V^{v}_T \\
\end{pmatrix}
,
\label{matrho5/2+}
\end{align}

\begin{align}
   V^{v}_{7/2^-} = &
\begin{pmatrix}
 V^{v\,\prime}_C&\displaystyle -3\sqrt{\frac{3}{7}}V^{v}_T&2\sqrt{3}V^{v}_C&\displaystyle \sqrt{\frac{15}{7}}V^{v}_T\\
\displaystyle  -3\sqrt{\frac{3}{7}}V^{v}_T&\displaystyle V^{v\,\prime}_C+2V^{v}_C+\frac{4}{7}V^{v}_T&\displaystyle -\frac{3}{\sqrt{7}}V^{v}_T&\displaystyle -\frac{6}{7}\sqrt{5}V^{v}_T\\
 2\sqrt{3}V^{v}_C&\displaystyle -\frac{3}{\sqrt{7}}V^{v}_T&V^{v\,\prime}_C-4V^{v}_C&\displaystyle \sqrt{\frac{5}{7}}V^{v}_T\\
\displaystyle  \sqrt{\frac{15}{7}}V^{v}_T&\displaystyle -\frac{6}{7}\sqrt{5}V^{v}_T&\displaystyle \sqrt{\frac{5}{7}}V^{v}_T&\displaystyle V^{v\,\prime}_C+2V^{v}_C-\frac{4}{7}V^{v}_T\\
\end{pmatrix}
,
\label{matrho7/2-}
\end{align}

\begin{align}
    V^{v}_{7/2^+} = &
\begin{pmatrix}
 V^{v\,\prime}_C&2\sqrt{3}V^{v}_C&-V^{v}_T&\sqrt{5}V^{v}_T\\
 2\sqrt{3}V^{v}_C&V^{v\,\prime}_C-4V^{v}_C&\displaystyle -\frac{1}{\sqrt{3}}V^{v}_T&\displaystyle \sqrt{\frac{5}{3}}V^{v}_T\\
 -V^{v}_T&\displaystyle -\frac{1}{\sqrt{3}}V^{v}_T&\displaystyle V^{v\,\prime}_C+2V^{v}_C-\frac{4}{3}V^{v}_T&\displaystyle -\frac{2}{3}\sqrt{5}V^{v}_T\\
 \sqrt{5}V^{v}_T&\displaystyle \sqrt{\frac{5}{3}}V^{v}_T&\displaystyle -\frac{2}{3}\sqrt{5}V^{v}_T&\displaystyle V^{v\,\prime}_C+2V^{v}_C+\frac{4}{3}V^{v}_T\\
\end{pmatrix}
,
\label{matrho7/2+}
\end{align}
where $V^{v\,\prime}_C$, $V^{v}_C$ and $V^{v}_T$ are defined as
\begin{align}
& V^{\rho\,\prime}_C= \frac{g_V g_{\rho NN} \beta
 }{\sqrt{2}m_{\rho}^2}C_{m_\rho}\vec{\tau}_{P} \cdot \vec{\tau}_N \, ,\\
& V^{\rho}_C= \frac{g_V g_{\rho NN} \lambda (1 + \kappa)}{
 \sqrt{2}m_N}\frac{1}{3}C_{m_\rho}\vec{\tau}_{P} \cdot \vec{\tau}_N \, , \\
& V^{\rho}_T= \frac{g_V g_{\rho NN} \lambda (1 + \kappa)}{
 \sqrt{2}m_N}\frac{1}{3}T_{m_\rho}\vec{\tau}_{P} \cdot \vec{\tau}_N \, , \\
& V^{\omega\,\prime}_C=\frac{g_V g_{\omega NN} \beta
 }{\sqrt{2}m_{\omega}^2}C_{m_\omega} \, , \\
& V^{\omega}_C= \frac{g_V g_{\omega NN}
 \lambda}{\sqrt{2}m_N}\frac{1}{3}C_{m_\omega} \, , \\
& V^{\omega}_T= \frac{g_V g_{\omega NN} \lambda}{
 \sqrt{2}m_N}\frac{1}{3}T_{m_\omega} \, .
\end{align}

Finally, the kinetic terms are given by
\begin{align}
 K_{1/2^-} &= \mbox{diag} \left( -\frac{1}{2 \tilde{m}_P}\bigtriangleup_0 ,
- \frac{1}{2\tilde{m}_{P^*}} \bigtriangleup_0 +\Delta m_{PP^*},
-\frac{1}{2\tilde{m}_{P^*}} \bigtriangleup_2 +\Delta m_{PP^*} \right)\,
 , \label{K1/2-} 
\end{align}
\begin{align}
 K_{1/2^+} &= \mbox{diag} \left( -\frac{1}{2 \tilde{m}_P}\bigtriangleup_1 ,
- \frac{1}{2\tilde{m}_{P^*}} \bigtriangleup_1 +\Delta m_{PP^*},
-\frac{1}{2\tilde{m}_{P^*}} \bigtriangleup_1 +\Delta m_{PP^*} \right)\,
 , \label{K1/2+} 
\end{align}
\begin{align}
K_{3/2^-} &= \mbox{diag} \left( -\frac{1}{2 \tilde{m}_P}\bigtriangleup_2 ,
- \frac{1}{2\tilde{m}_{P^*}} \bigtriangleup_0 +\Delta m_{PP^*},
-\frac{1}{2\tilde{m}_{P^*}} \bigtriangleup_2 +\Delta m_{PP^*}
\right. , \notag \\
& \quad \left.-\frac{1}{2\tilde{m}_{P^*}} \bigtriangleup_2 +\Delta
 m_{PP^*} \right) \, , \label{K3/2-} 
\end{align}
\begin{align}
K_{3/2^+} &= \mbox{diag} \left( -\frac{1}{2 \tilde{m}_P}\bigtriangleup_1 ,
- \frac{1}{2\tilde{m}_{P^*}} \bigtriangleup_1 +\Delta m_{PP^*},
-\frac{1}{2\tilde{m}_{P^*}} \bigtriangleup_1 +\Delta m_{PP^*}
\right. , \notag \\
& \quad \left.-\frac{1}{2\tilde{m}_{P^*}} \bigtriangleup_3 +\Delta
 m_{PP^*} \right) \, , \label{K3/2+} 
\end{align}
\begin{align}
K_{5/2^-} &= \mbox{diag} \left( -\frac{1}{2 \tilde{m}_P}\bigtriangleup_2 ,
- \frac{1}{2\tilde{m}_{P^*}} \bigtriangleup_2 +\Delta m_{PP^*},
-\frac{1}{2\tilde{m}_{P^*}} \bigtriangleup_2 +\Delta m_{PP^*}
\right. , \notag \\
& \quad \left.-\frac{1}{2\tilde{m}_{P^*}} \bigtriangleup_4 +\Delta
 m_{PP^*} \right) \, , \label{K5/2-} 
\end{align}
\begin{align}
K_{5/2^+} &= \mbox{diag} \left( -\frac{1}{2 \tilde{m}_P}\bigtriangleup_3 ,
- \frac{1}{2\tilde{m}_{P^*}} \bigtriangleup_1 +\Delta m_{PP^*},
-\frac{1}{2\tilde{m}_{P^*}} \bigtriangleup_3 +\Delta m_{PP^*}
\right. , \notag \\
& \quad \left.-\frac{1}{2\tilde{m}_{P^*}} \bigtriangleup_3 +\Delta
 m_{PP^*} \right) \, , \label{K5/2+} 
\end{align}
\begin{align}
K_{7/2^-} &= \mbox{diag} \left( -\frac{1}{2 \tilde{m}_P}\bigtriangleup_4 ,
- \frac{1}{2\tilde{m}_{P^*}} \bigtriangleup_2 +\Delta m_{PP^*},
-\frac{1}{2\tilde{m}_{P^*}} \bigtriangleup_4 +\Delta m_{PP^*}
\right. , \notag \\
& \quad \left.-\frac{1}{2\tilde{m}_{P^*}} \bigtriangleup_4 +\Delta
 m_{PP^*} \right) \, , \label{K7/2-} 
\end{align}
\begin{align}
K_{7/2^+} &= \mbox{diag} \left( -\frac{1}{2 \tilde{m}_P}\bigtriangleup_3 ,
- \frac{1}{2\tilde{m}_{P^*}} \bigtriangleup_3 +\Delta m_{PP^*},
-\frac{1}{2\tilde{m}_{P^*}} \bigtriangleup_3 +\Delta m_{PP^*}
\right. , \notag \\
& \quad \left.-\frac{1}{2\tilde{m}_{P^*}} \bigtriangleup_5 +\Delta
 m_{PP^*} \right) \, , \label{K7/2+}
\end{align}
where 
$\bigtriangleup_l = \partial^2 / \partial r^2 + (2/r)\partial / \partial
r  - l(l+1)/r^2$,
$\tilde{m}_{P{^{(\ast)}}}
= m_N m_{P{^{(\ast)}}}/(m_N +m_{P{^{(\ast)}}}),$ and $\Delta m_{PP^*} = m_{P^*} -m_P$.
The total Hamiltonian is then given by $H_{J^P} = K_{J^P} + V_{J^P}$.


\end{document}